# Brush in the bath of active particles: anomalous stretching of chains and distribution of particles

Hui-shu Li[1], Bo-kai Zhang[2,1], Jian Li[3,1], Wen-de Tian[1,4,*] and Kang Chen[1,4,*]

[1]Center for Soft Condensed Matter Physics & Interdisciplinary Research, College of Physics, Optoelectronics and Energy, Soochow University, Suzhou 215006, China.

[2]National Laboratory of Solid State Microstructures and Department of Physics, Nanjing University, Nanjing 210093, China

[3]Department of Physics, Nanjing Normal University, Nanjing 210023, China

[4]Kavli Institute for Theoretical Physics China, CAS, Beijing 100190, China

[*]Authors to whom Correspondence should be addressed. Electronic mail: tianwende@suda.edu.cn (W.d.T.); kangchen@suda.edu.cn (K.C.)

## Abstract

The interaction between polymer brush and colloidal particles has been intensively studied in the last two decades. Here we consider a flat chain-grafted substrate immersed in a bath of active particles. Simulations show that an increase in the self-propelling force causes an increase in the number of particles that penetrate into the brush. Anomalously, the particle density inside the main body of the brush eventually becomes higher than that outside the brush at very large self-propelling force. The grafted chains are further stretched due to the steric repulsion from the intruded particles. Upon the increase of the self-propelling force, distinct stretching behaviors of chains were observed for low and high grafting densities. Surprisingly, we found a weak descent of the end-to-end distance of chains for high grafting density and very large force which is reminiscent of the compression effect of a chain in the active bath.

**Introduction**

Polymer brush can be formed by a layer of polymers tethered with one end to a substrate. Such structure is widely used to modify or improve the properties of a surface such as lubrication,[1-3] chemical affinity [4,5] and biocompatibility.[6,7] In colloidal suspensions, it has been used to tune the interactions between the colloidal particles and thereby control their assembly behavior. For example, densely grafted chains can generate effective steric repulsions between colloidal particles and thereby prevent flocculation;[8] grafted-chain-driven associations can direct the assembly of nanoparticles into colloidal polymers or networks.[9] In biosystems, the lipid tails in the membrane form a typical structure of brush; arrays of cilia are also of brush-like structure but of much larger size (micrometer) than the conventional molecular brush.[10,11]

Many efforts have been made to reveal and understand the equilibrium properties of polymer brush.[12-16] In good solvent and for the grafting density high above the overlap value, the excluded-volume interaction between monomers stretches the chains in the direction perpendicular to the surface. In contrast to the sub-linear scaling of the end-to-end distance $R_e$ with chain length $N$ ($R_e \propto N^\nu$ with Flory exponent $\nu \approx 0.588$) for a free polymer in solution,[17] a linear scaling of brush height $h$ with chain length $N$ ($h \propto N$) is predicted for polymer brush.[13] More scaling behaviors of polymer brush in conditions such as marginal solvents, Theta solvents, and poor solvents, are summarized in a recent review.[18] The typical density profile of a brush in good solvent consists of three distinct regimes: i) oscillation regime close to the surface (layering effect), ii) relative flat regime in the middle and iii) smooth decay regime at the top. The profile varies quantitatively with the effective monomer-monomer interaction, chain length and grafting density, etc.

The interplay between polymer brush and nanoparticles is an important aspect due to the applications in nanomaterials[19-22] and the relevance to biotechnology such as protein adsorption, cell adhesion and drug encapsulation, etc.[6, 23-28] Theoretical and simulation studies have shown that the solubility, the size and the shape of particles greatly influence their spatial organization on the brush.[29-42] For example, for soluble spherical nanoparticles, there is an upper threshold size beyond which particles cannot penetrate into the brush and a lower threshold size below which particles can completely penetrate into the brush; and for the size in between particles can partly penetrate into the brush with a thickness proportional to the brush height and inversely proportional to the particle volume.[40, 43]

Recently, there is an increasing interest in the behaviors of a chain immersed in a bath of active particles.[44-46] The passive motion of the chain in the active bath receives additional non-thermal fluctuation from the self-propelling motion of particles through collisions, which consequently leads to anomalous nonequilibrium static and dynamic properties of the chain. It was shown that, for a flexible chain, the conventional Flory scaling between $R_e$ and $N$ is no longer hold when the chain length is not very large.[44] Instead, a nonuniversal behavior was found between $R_e$ and $N$ under various activities. Detailed observation on the snapshots showed that different kinds of active motion of particles near the chain could compress or stretch it. Rigidity of a semiflexible chain could lead to more complex and richer phenomena in such nonequilibrium system. For example, the hairpin configurations were found to be metastable under certain strengths of bath activity.[45]

In this paper, we studied the system of a chain-grafted substrate immersed in a bath of active particles. Our focus is on the variation of the particle distribution and chain configuration with the strength of the driving force on the particles. We found the self-propelling force drives more particles into the body of brush. And eventually, the particle density in the bulk of the brush becomes larger than that out of the brush at very large active force. Meanwhile, the configuration of the grafted chains varies due to the particle-chain collisions. Our data show distinct stretching behaviors of the chains for the low and high grafting densities.

**Model and Simulation Methods**

In our simulation, a grafted chain is modeled as a sequence of bond-connected beads with one of its ends attached to a flat substrate. Each active particle is treated as a self-propelled sphere driven by a force $F$ along the direction $\hat{\mu}$ which orientates under thermal fluctuation. The schematic of the system is shown in Fig. 1.

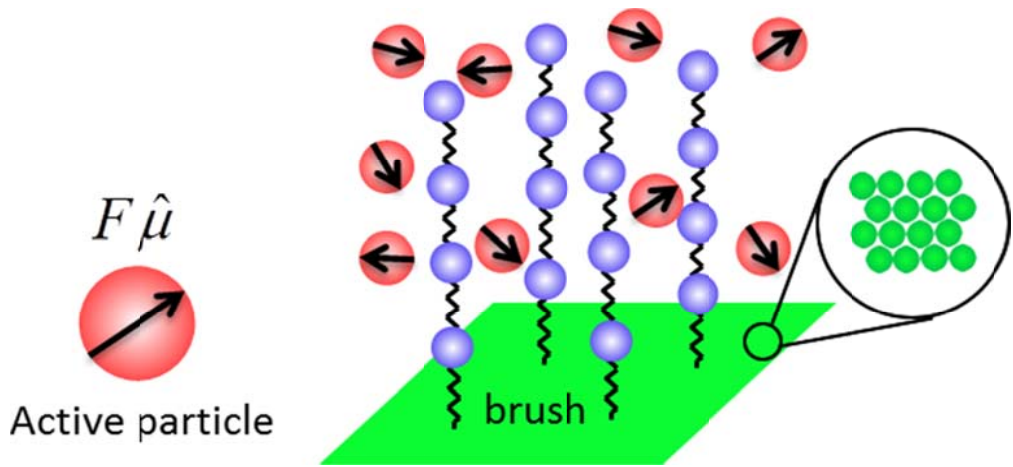

**Fig.1.** A schematic of the system. Bead-spring chains are grafted on a planar substrate, which is composed of one-layer hexagonally-packed beads. The active particles are modeled as self-

propelled spheres, with the driving force along the direction, $\hat{u}$. To prevent the escape of particles from the top, a smooth repulsive wall is put on the top of the simulation box (not shown).

The purely repulsive Weeks-Chandler-Andersen (WCA) potential is adopted for the non-bonded interactions between all beads and active particles,[47]

$$U_{WCA}(r) = \begin{cases} 4\varepsilon\left[\left(\dfrac{\sigma}{r}\right)^{12} - \left(\dfrac{\sigma}{r}\right)^{6}\right] + \varepsilon & r < r_c = \sqrt[6]{2}\sigma \\ 0 & r > r_c \end{cases} \quad (1)$$

Here, we assume all the beads and active particles are of the same diameter, $\sigma$. $\varepsilon$ is the interaction strength. The bonded interaction between successive beads is described by the finitely extensible non-linear elastic (FENE) potential,[48]

$$U_{FENE}(r_{i,i+1}) = -\frac{1}{2}kR_0^2 \ln\left[1 - (r_{i,i+1}/R_0)^2\right] \quad (2)$$

where $r_{i,i+1} = |\mathbf{r}_i - \mathbf{r}_{i+1}|$, and $\mathbf{r}_i$, $\mathbf{r}_{i+1}$ are position vectors of two successive beads along a chain. $k$ is the spring constant, $R_0$ the maximum length of the bond. We set $R_0 = 1.5\sigma$, $k = 30\varepsilon/\sigma^2$.[48, 49] The chains are randomly grafted onto the substrate. The substrate is mimicked by a layer of hexagonally closely-packed spherical particles. The WCA repulsive potential induced by the substrate particles prevents the components in the system from passing through the substrate.

We use the Langevin dynamics to describe the Brownian motion of the chain beads,

$$m\ddot{\mathbf{r}}_j^i = -\frac{\partial U_j^i}{\partial \mathbf{r}_j^i} - \gamma \dot{\mathbf{r}}_j^i + \boldsymbol{\eta}_j^i(t) \quad (3)$$

where $\mathbf{r}_j^i$ is the coordinate of the $j$th bead of the $i$th chain, $m$ is the mass of a bead or an active particle. The first term on the right hand side (rhs) of the equation denotes the deterministic force acting on the bead by other explicit entities in the system. And the potential $U_j^i$ is composed of both non-bonded WCA repulsion and bonded FENE potential. The second and third terms on the rhs are damping viscous term and Gaussian white noise accounting for the roles of the implicit solvent. The noise term $\eta_j^i(t)$ satisfies the fluctuation-dissipation theorem, $<\eta_{j,\alpha}^i(t)\eta_{l,\beta}^k(t')> = 2D_0\delta_{ik}\delta_{jl}\delta_{\alpha\beta}\delta(t-t')$, where $\alpha$ and $\beta$ denote components of Cartesian coordinates. $D_0$ is the translational diffusion constant of a single bead. The motion of the active particles follows the equations

$$m\ddot{\boldsymbol{r}}_i = -\frac{\partial U_i}{\partial \boldsymbol{r}_i} - \gamma\dot{\boldsymbol{r}}_i + F\hat{\boldsymbol{\mu}}_i(t) + \boldsymbol{\eta}_i(t) \tag{4}$$

$$\partial_t \hat{\boldsymbol{u}}_i(t) = \boldsymbol{\zeta}_i(t) \times \hat{\boldsymbol{u}}_i(t) \tag{5}$$

where $\boldsymbol{r}_i$ is the position of the $i$th active particle and $\hat{\boldsymbol{\mu}}_i = (\sin\theta\cos\varphi, \sin\theta\sin\varphi, \cos\theta)$ the unit vector along the direction of the self-propelling force $F$ on it. $\boldsymbol{\eta}_i(t)$ is Gaussian white noise as in eq. (3), while $\boldsymbol{\zeta}_i(t)$ is also a Gaussian noise, which satisfy the relation, $\langle \zeta_{i,\alpha}(t)\zeta_{i,\beta}(t')\rangle = 2D_r\delta_{ij}\delta_{\alpha\beta}\delta(t-t')$. $D_r = 3D_0/\sigma^2$ is the rotational diffusion constant.

In our system, there are $N_s$ active particles and $n$ grafted chains, each of which contains $N$ beads. The total number of beads is $N_b = nN$. The grafting density is defined as $g_r = n/A$, where $A = 25\sigma \times 25\sigma$ is the area of the substrate. We set $N=70$ and studied the cases of two grafting densities, $g_r = 0.2$ and 0.4. The ratio of the number of active particles to the total number of beads is fixed, $N_s/N_b = 0.2$. The number density of active particles in all of our simulations is sufficient small to avoid spontaneous phase separation.[50] We used LAMMPS software to perform our simulations.[51] Periodic boundary condition is adopted in the xy (horizontal) directions. The substrate is put at $z=0$. A fixed boundary of a smooth wall with

WCA repulsive potential is also put high above the brush to prevent the active particles from moving too far away. Reduced units are used in the simulation by setting $m=1$, $\varepsilon=1$ and $\sigma=1$. The corresponding unit time, $\tau=\sqrt{m\sigma^2/\varepsilon}$. We choose the reduced temperature, $k_BT=1.2$, and the friction coefficient $\gamma=10$. For every case, it was run by a total time of $10^5\tau$ with a time step, $\Delta t=0.001\tau$, and only the data of the last $4\times10^4\tau$ were used for analysis.

**Results and Discussion**

One concerned question for this system is how the structure of the brush is affected by the presence of active particles. The self-propelled motion of these particles induces additional non-equilibrium fluctuations to the grafted chains. Fig. 2(a) and (b) show the density profiles, $\rho(z)$, of the brush at grafting densities $g_r=0.2$ and 0.4, respectively. The self-propelling force varies from 0 to 80. As generally found in brushes,[12, 18] density of beads fluctuates close to the grafting surface. The fluctuating peaks are more prominent in the case of high grafting density, implying the formation of more manifested layer structures of beads near the substrate. At large *z*, the density profiles of beads are smooth and gradually decay to 0. Apparently, the profiles quantitatively change with the driving force *F* on the active particles. The insets of Fig. 2(a) and (b) zoom in the density profiles near the top of the brush. Within the scope of forces we have examined, the brush profiles of low grafting density extend to larger *z* with increasing active force, monotonically. While, non-monotonic variation was found for the brush of high grafting density, i.e. the profile first extends gradually at low active force then contracts a little bit at high active force. These behaviors can be quantified by calculating the density-averaged height of the brush defined as $h=\int\rho(z)zdz/\int\rho(z)dz=\frac{1}{nN}\sum_{i=1}^{n}\sum_{j=1}^{N}z_j^i$. Fig. 2(c) shows the reduced height of the brush as a function of active force. $h_0$ is the height of the brush when $F=0$. The heights at both the low and high grafting densities initially increase rapidly with increasing active force. The height of $g_r = 0.2$ reaches maximum around $F = 40$ and becomes nearly constant for higher $F$. Contrastingly, the brush height of $g_r = 0.4$ shows an evident decrease for $F > 50$. The extension of brush can also be analyzed based on the averaged end-to-end distance $R_e$ of the grafted chains. $R_e^\perp = \sqrt{\frac{1}{n}\sum_{i=1}^{n}\langle z_N^{i\,2}\rangle}$ and $R_e^\| = \sqrt{\frac{1}{n}\sum_{i=1}^{n}\langle (x_N^i - x_0^i)^2 + (y_N^i - y_0^i)^2 \rangle}$ are the vertical and horizontal components of the end-to-end distance which reflect the extension of chains along the $z$ and $xy$ directions, respectively. The $R_e^\perp$ and $R_e^\|$ scaled by their values at $F = 0$ are shown in Fig. 2(d). Different from the height of brush, $R_e^\perp$ for $g_r = 0.2$ increases gradually with active force without an apparent plateau after $F > 40$. For $g_r = 0.4$, $R_e^\perp$ shows a similar non-monotonic dependence on active force as the height of the brush. Trivially, the curves of $R_e^\|$ show opposite trends in contrast to those of $R_e^\perp$.

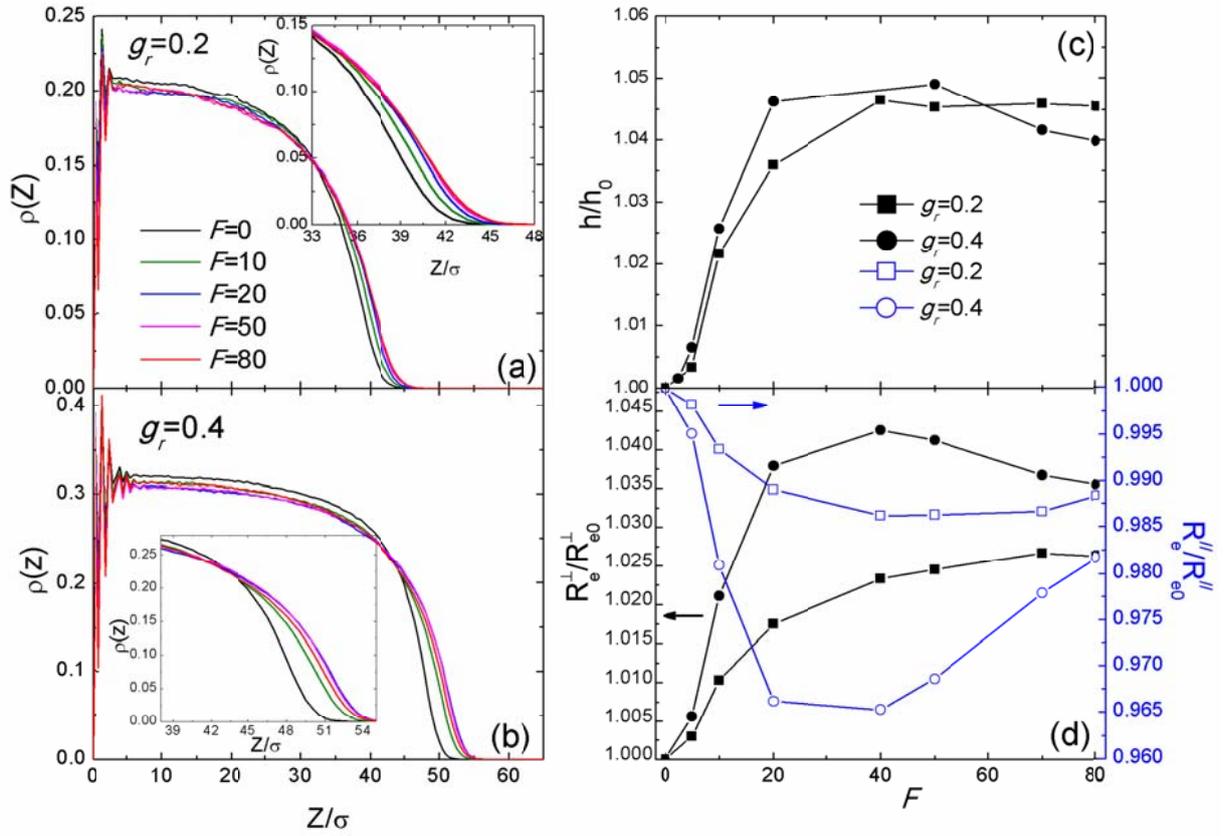

**Fig.2** The number density of chain beads, $\rho(z)$, as a function of the distance from the grafting substrate for different self-propelling forces at two grafting densities (a) $g_r = 0.2$, (b) $g_r = 0.4$; The insets of (a) and (b) zoom in the density profiles of chain beads near the top of brush. (c) The reduced statistical averaged thickness, $h/h_0$, of the brush as a function of self-propelling force. $h_0$ is the thickness of brush when the particles are non-active, i.e. $F = 0$. (d) The vertical ($R_e^\perp$, black lines), and horizontal ($R_e^\parallel$, blue lines) components of the averaged end-to-end distance, reduced by their values at $F = 0$, as functions of self-propelling force.

Before we try to understand the behaviors of the brush, we need to know how the distribution of active particles changes with the self-propelling force. Fig. 3 shows the density profiles, $\rho_p(z)$, of active particles for two grafting densities and various active forces. Clear peaks emerge near $z=0$ indicating the formation of a high-density layer of particles near the substrate. The peak or the density of the layer of particles increases dramatically with increasing force. For both grafting densities, the curves of different forces intersects near the top of the brush (comparing Fig. 3 with Fig. 2(a) and (b)). Apparently, the density profiles are naturally divided into two parts, i.e., roughly, the part to the left (right) of the intersection represents the particles inside (outside) the brush. As anticipated, the increase of active force on the particles effectively enhances their penetrability, and therefore, more and more particles enter the brush. A novel finding is that the particle density inside the main body of the brush (the flat part of the curve) becomes larger than that outside the brush when the active force is large ($F > 20$). This anomalous density distribution is more prominent in the case of high grafting density, $g_r = 0.4$. A possible explanation for this phenomenon is as follows. At low active force, the transportation of particles is diffusion-controlled. The excluded-volume interactions between grafted chains and particles lead the formation of normal (positive) particle density gradient from the inside to the outside of the brush. But, the active contribution becomes dominated in the motion of particles when the force is very large. The active motion of particles inside the brush is frustrated by the presence of grafted chains. To balance the mobility difference between particles inside and outside the brush, an anomalous (negative) density gradient thus takes place. This frustration effect is enhanced in the case of dense brush, which explains the observed more prominent anomalous density difference for $g_r = 0.4$.

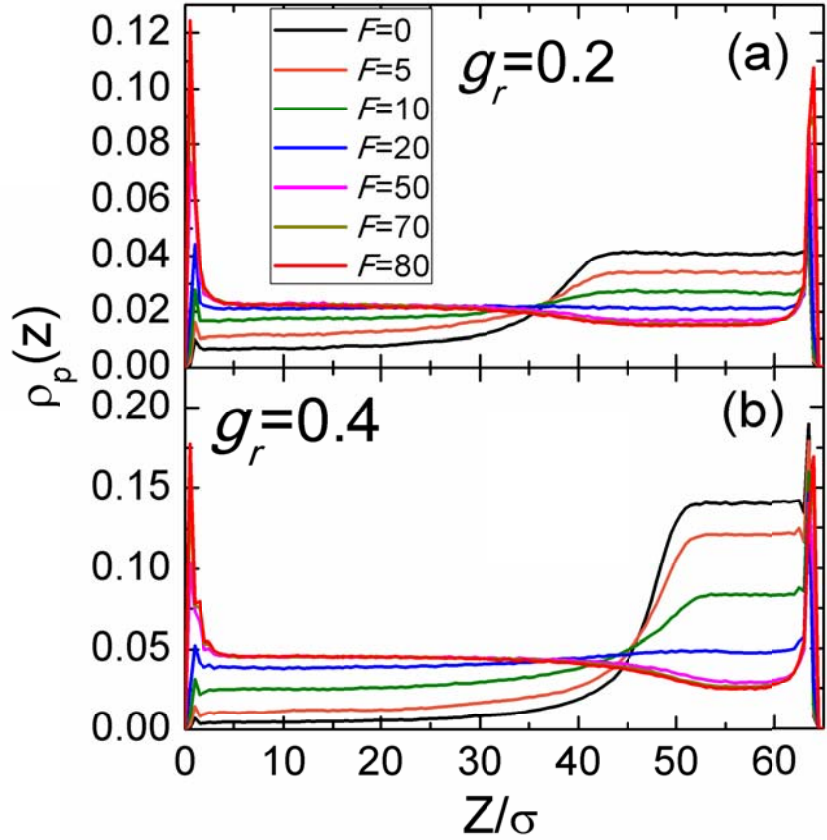

**Fig. 3.** The number density distribution of active particles, $\rho_p(z)$, for various self-propelling forces at grafting densities, (a) $g_r = 0.2$, and (b) $g_r = 0.4$.

To quantitatively describe the variation of the particle distribution, we counted statistically the number of particles that penetrate into the brush, $N_s^{inside}$. We chose $2h$ as the upper border of the brush, i.e. the particles at $z < 2h$ are considered inside the brush. The ratio of $N_s^{inside}/N_s$, i.e. the proportion of inside particles, is shown in Fig. 4. This ratio for both grafting densities rises significantly with increasing active force when $F < 20$, and then gradually approaches a plateau which is consistent with the nearly superposed density profiles (flat part) for active forces above

20. For non-active particles ($F=0$), the ratio for $g_r=0.4$ is smaller than that for $g_r=0.2$ due to stronger excluded-volume interaction potential between chain beads and particles in denser brush. But the ratio for $g_r=0.4$ rises faster with increasing active force. We also kept record of the effective temperature, $T_{brush}$, of the brush based on the thermal motion of chain beads (Fig. 4). This effective temperature could deviate from the ambient temperature due to the additional disturbance from the non-equilibrium active motion of particles. It shows that the effective temperature starts from ambient temperature $k_B T =1.2$ at small active force and then rises monotonically. Both the increase of the number of particles inside the brush and the enhanced active motion of individual particles could contribute to the rise of the effective temperature. For $g_r=0.4$, the denser brush and more number of particles inside the brush lead to more frequent collisions between active particles and chain beads, and consequently the faster rising of the effective temperature.

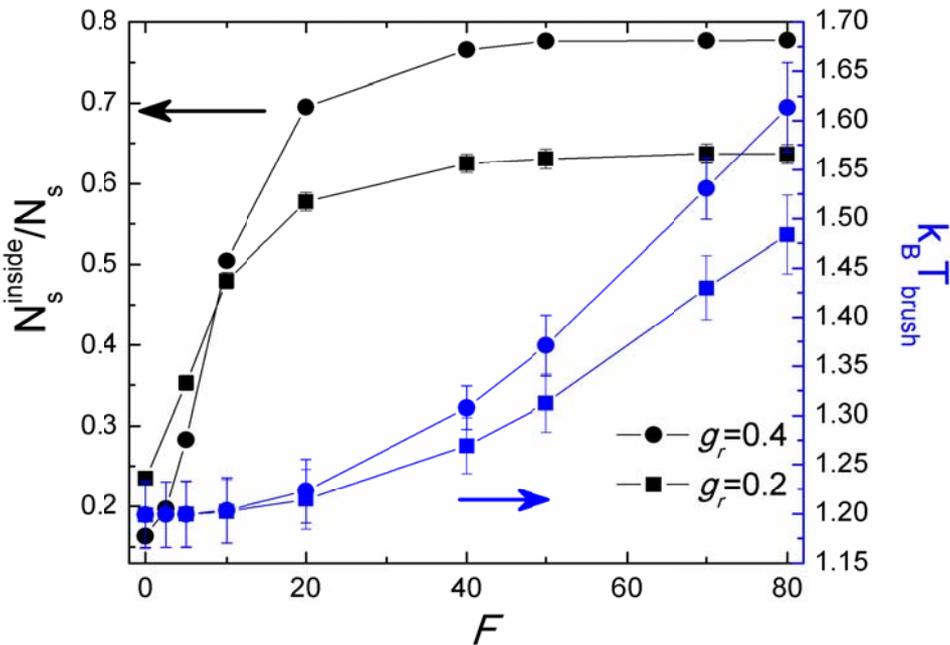

**Fig. 4.** The proportion of active particles inside the brush as a function of self-propelling force (left vertical axis and black curves). $N_s^{inside}$ is the number of active particles inside the brush, and $N_s$ the total number of particles in the system. Also shown is the effective temperature of the brush beads, $k_B T_{brush}$ (right vertical axis and blue curves).

Knowing the distributions of particles, we come back to further analyze the variation of configurations of the grafted chains. We wanted to know how the different parts of the grafted chains are stretched. To this end, we picked three blocks of equal number of beads in each grafted chain: starting from the grafting end, the ground block contains the beads from 1 to 20, the middle block from 25 to 45, and the top block from 50 to 70. We compared the variation of $R_e^\perp$'s of these three blocks in Fig. 5, which roughly reflect the extension of brush chains near the substrate, in the middle and at the top. For F<20, $R_e^\perp$'s of all three blocks for both grafting densities increase rapidly with increasing active force. This is ascribed to the evident increase of the particle density inside the brush (see Fig. 3 and 4), which stretch the blocks by the excluded-volume interaction. Situation is complicated at large active force, under which the curves of $R_e^\perp$ for low and high grafting densities show distinct dependence on the active force. Look at the particle distributions inside the brush for forces above 20 in Fig. 3, the particle density increases significantly near the substrate but saturates in the main body of the brush with increasing active force. In the view of the excluded-volume interaction, one would expect the curve of $R_e^\perp$ of the ground block monotonically increases but $R_e^\perp$'s of middle and top blocks should be nearly unchanged at large active force. This excluded-volume argument roughly holds for the low grafting density, $g_r = 0.2$ (Fig. 5). And, the monotonic increase of the whole-chain $R_e^\perp$ of $g_r = 0.2$ in Fig. 2(d) at large active force is contributed mostly from the monotonic stretching of the ground block. However, for the high grafting density, $g_r = 0.4$, $R_e^\perp$ of the ground block does not increase at large force; instead, $R_e^\perp$'s of all three blocks show slight but clear descent which result in the non-monotonic behavior of $R_e^\perp$ of the whole chain in Fig. 2(d). A possible explanation for the ground block not stretching further when the particles become denser near the substrate is that the ground block for high grafting density is already stretched a lot at $F$ above 20 and only very few beads close to the substrate are influenced by the enhanced excluded-volume interactions from the denser particle layer. The novel descent behavior of $R_e^\perp$'s is probably related to the phenomenon of conformational change of polymer chain in the bath of active particles.[44] Active particles move along the chain can stretch it but collisions perpendicular to the chain may cause the compression effect, i.e. the decrease of $R_e$. Note that, in our 3D system, these stretching and compression effects of chains due to the active motion of particles should be weaker than in 2D system. But, they may be manifested in the case of dense 3D brush. For $g_r = 0.4$, the chains are highly stretched and the compression effect may dominates the interplay between the motion of active particles and the change of chain conformation. This compression effect is enhanced at large active force and accounts for the descent of $R_e^\perp$'s in Fig. 5.

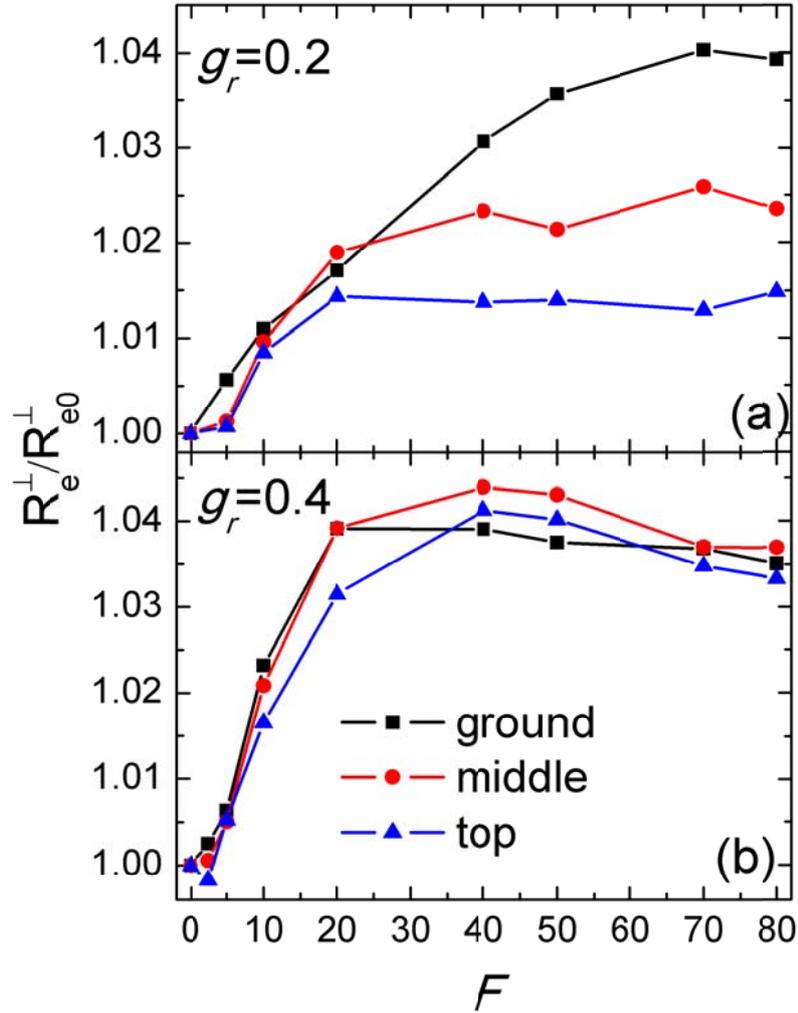

**Fig. 5.** The vertical components of the end-to-end distances of the ground, middle and top blocks of grafted chains. The grafting densities are (a) $g_r = 0.2$, and (b) $g_r = 0.4$. $R_{e0}^{\perp}$ represents the corresponding value in the case of zero active force.

**Conclusions**

In this paper, we performed 3D Brownian dynamics simulations to study the chain-grafted substrate immersed in the bath of active particles. We are concerned with the impact of the non-thermal active motions of particles on their distribution and the configuration of grafted chains.

To this end, we examined the density profiles of the brush and particles, the end-to-end distances of the grafted chains and three blocks, which are near the substrate, in the middle and at the top of the brush, respectively. We found the self-propelling force can facilitate the penetration of particles into the brush. The particle density in the bulk of the brush saturates at large active force and counterintuitively, becomes eventually larger than the particle density outside the brush. This anomalous particle density distribution may be ascribed to the lower active mobility of particles inside the brush. These inside particles stretch the grafted chains through the excluded-volume interaction. But the chains show distinct stretching behaviors for low and high grafting densities. The excluded-volume argument can well interpret the extensions of the brush or the blocks of the grafted chains for low grafting density, $g_r = 0.2$. However, surprising slight but clear descents of the brush height and the end-to-end distances of the three blocks and the whole chain at very large active force were observed for high grafting density, $g_r = 0.4$. We believe this novel phenomenon is probably correlated with the unusual conformational change of polymer chain in the bath of active particles. This work provides some preliminary and interesting observations on the interplay between the passive grafted substrate and active elements. Further explorations on the microscopic mechanisms and systematic studies on the nonequilibrium phenonmena in analogous systems with active agents of various shapes and sizes, etc and for various grafting conditions are anticipated in the future.

**Acknowledgements**

This work was supported by the National Basic Research Program of China (973 Program) No. 2012CB821500 and the National Natural Science Foundation of China (NSFC) Nos. 21374073, 11074180, 21574096 and 21474074.